# Magnetodielectric Effects from Spin Fluctuations in Isostructural Ferromagnetic and Antiferromagnetic Systems

G. Lawes[1], A.P. Ramirez[1], C.M. Varma[2], M. A. Subramanian[3]

[1]*Los Alamos National Laboratory, Los Alamos NM, 87545*
[2]*Bell Laboratories, Lucent Technologies, 600 Mountain Avenue, Murray Hill, NJ 07974*
[3]*DuPont Central Research and Development, Experimental Station, Wilmington, DE 19880*

We report on the effects of spin fluctuations, magnetic ordering, and external magnetic field on the dielectric constant of the ferromagnet $SeCuO_3$, and the antiferromagnet $TeCuO_3$. A model based on the coupling between uniform polarization and the q-dependent spin-spin correlation function is presented to explain the different behaviors for these isostructural compounds. The large magnetocapacitance near the transition temperature in the ferromagnet $SeCuO_3$ suggests routes to enhancing the magnetodielectric response for practical applications.

The behavior of systems with strongly coupled magnetic and electronic degrees of freedom provides both challenges for many-body theory as well as new phenomena for possible applications. One manifestation of such coupling in the itinerant limit is the interplay between ferromagnetism and charge order in the colossal magnetoresistance manganites [1,2]. In the limit of localized charge, effects of strong coupling are more subtle, and manifested chiefly through a magnetocapacitive (MC) response, which can take several different forms [3,4,5]. The present strong interest in dielectric device properties has motivated the search for so-called multiferroic materials [6,7] – compounds possessing both ferromagnetic and ferroelectric order arising from different atomic constituents. After the first experimental realization of magnetoelectric coupling in $Cr_2O_3$ [8], similar effects have been observed in several other materials including $Gd_2CuO_4$ [9], $YMnO_3$ [10], $EuTiO_3$ [11], and $BiMnO_3$ [12].

In this work we compare the magnetodielectric [13] (MD) response in a ferromagnetic insulator and an isostructural antiferromagnet. We investigate the effect of ferromagnetic (FM) and antiferromagnetic (AF) magnetic correlations on the dielectric constant, ε, by measuring the sample capacitance as a function of temperature and magnetic field. We find differences between the FM and AF samples in the temperature dependence of ε both in the paramagnetic and the magnetically ordered regimes. Moreover, the magnetic field dependence is quite different near their respective transition temperatures.

The structure of $SeCuO_3$ and $TeCuO_3$ is that of a distorted perovskite with the small $Se^{4+}$ or $Te^{4+}$ ions on the A-cation sites producing a Cu-O-Cu bond angle of $\alpha_{Cu\text{-}O} = 121^o$ (structure shown in the inset of fig. 1)[13]. For $SeCuO_3$, the $Cu^{2+}$ ions undergo a ferromagnetic (FM) phase transition at $T_c = 25$ K, with a saturation magnetization of 0.7 $\mu_B$ per Cu ion. Both the FM transition as well as the reduced moment have been understood to arise from the dependence of the superexchange interaction on $\alpha_{Cu\text{-}O} = 127.1^o$, which for $SeCuO_3$ sits on the FM side of the Goodenough-Kanemori value ($\alpha_{GK} = 127.5^o$). The isomorphic system $TeCuO_3$, on the other hand, exhibits antiferromagnetism which is reflected in a value for $\alpha_{Cu\text{-}O} = 130.5^o$, which sits just on the antiferromagnetic (AF) side of $\alpha_{GK}$[14]. Both materials are good insulators, and this makes them uniquely well-suited to study effects that depend on the sign of the superexchange interaction without complications brought about by differences in other materials factors (e.g. range of interaction, single-ion anisotropy, disorder).

The samples used in the study were made by solid state reaction at 700 C under 60 kbar pressure using high purity $SeO_2$, $TeO_2$ and CuO starting materials. The purity of the phases was checked by X-ray powder diffraction and the details were given elsewhere [14]. We measured the magnetization of $SeCuO_3$ and $TeCuO_3$ using a SQUID magnetometer, both as function of temperature at fixed field, and as a function of applied magnetic field at fixed temperature. The inverse magnetization of these samples versus temperature, and the magnetization of $SeCuO_3$ at an applied field of 1 kOe, are shown as the upper plot in figure 1. The $SeCuO_3$ curves show the onset of a sharp ferromagnetic transition at $T_c = 25.0$ K, while the $TeCuO_3$ sample undergoes antiferromagnetic ordering at $T_N = 9$K; both are consistent with earlier measurements [14]. We prepared the samples for capacitance measurements by polishing opposite parallel faces and then depositing ~80 nm thick Au layers onto these smooth surfaces to serve as electrodes. Thin Pt wires were attached to the electrodes using silver epoxy. The samples were fixed to a glass plate on the probe using GE varnish to ensure mechanical stability. We measured the capacitance using an Agilent 4284A LCR meter. All dielectric measurements were done at a frequency of 1

MHz with an excitation of 1 V. Lower frequency measurements with different excitation voltages showed qualitatively similar behavior.

The dielectric response of these samples is shown in the lower plot in figure 1 as a function of temperature in the absence of an applied magnetic field. $SeCuO_3$ exhibits an almost temperature independent dielectric constant until just above $T_c$. $TeCuO_3$ on the other hand shows a pronounced increase in $\varepsilon$ as the sample is cooled. Both samples undergo a sharp drop in dielectric constant coincident with the onset of magnetic ordering, at 25 K for $SeCuO_3$ and 9K for $TeCuO_3$. The drop in the former is much larger than in the latter. Qualitatively similar features were observed in both the antiferromagnet $EuTiO_3$, in which the shift is attributed to the softening of an optical phonon mode at the AFM ordering transition [11], as well as in the insulating ferromagnet $BiMnO_3$ [12].

We have also investigated the dielectric response in $SeCuO_3$ as a function of temperature at fixed magnetic field (shown in Fig. (2)) and in both $SeCuO_3$ and $TeCuO_3$ as a function of magnetic field at fixed temperatures close to the magnetic ordering transition (Fig. (3)). These figures also include data on the magnetization of both the FM and AF samples taken under the same conditions as the dielectric measurements to investigate the effects of magnetic correlations on the dielectric constant.

The observations to be understood in Figs (1), (2), and (3) are: i) the temperature independence of $\varepsilon_0$ for both $SeCuO_3$ and $TeCuO_3$ at high temperatures (Fig. 1), ii) the rise in $\varepsilon$ for $TeCuO_3$ (AF) as the temperature is decreased towards to the transition temperature $T_N$, while for $SeCuO_3$ (FM) $\varepsilon$ remains almost constant as the temperature is reduced to $T_c$ (Fig.1), iii) the larger drop in $\varepsilon$ for $SeCuO_3$ (FM) in the ordered phase compared to that in $TeCuO_3$ (AF) (Fig. 1 and Fig. 2), iv) the sharp magnetic field dependence of ε near the ferromagnetic transition, while for the AF transition the dependence on H is smooth (Fig. 3).

All these observations are explained by a simple phenomenological model for the coupling of the uniform electric polarization P to the magnetization $M_q$ at wave-vector q. The lowest order free energy invariant considered is:

$$F = \frac{P^2}{2\varepsilon_0} - P \cdot E + P^2 \sum_q g(q) \left\langle M_q M_{-q} \right\rangle (T) \tag{1}$$

Here E is the applied electric field, $\varepsilon_0$ is the "bare" dielectric constant, g(q) the q-dependent coupling constant, and $\left\langle M_q M_{-q} \right\rangle$ is the thermal average of the instantaneous spin-spin correlation, which obeys the sum rule

$$\sum_q \left\langle M_q M_{-q} \right\rangle = N g^2 \mu_B^2 S(S+1)\,. \tag{2}$$

In addition to the term discussed in Eq. 1, there can also be couplings of the form $\lambda_{ijk}\left(P_i M_j \nabla_j M_k\right)$, which are important near domain walls. These couplings are unimportant in these polycrystalline samples examined here and are neglected.

Extremizing Eq. (1) with respect to the polarization P leads to:

$$P = \frac{E}{\frac{1}{\varepsilon_0} + 2\sum_q g(q)\left\langle M_q M_{-q} \right\rangle} \equiv \varepsilon E \tag{3}$$

so that the actual value of the dielectric constant ε is

$$\varepsilon = \frac{\varepsilon_0}{1 + 2\varepsilon_0 I(T)}\,, \quad I(T) = \sum_q g(q)\left\langle M_q M_{-q} \right\rangle (T)\,. \tag{4}$$

Given the sum-rule of Eq (2), it follows that the temperature dependence of ε depends on the relative q dependence of g(q) and of $\left\langle M_q M_{-q} \right\rangle$(T). At high temperatures where $\left\langle M_q M_{-q} \right\rangle$ is q and T independent, $\varepsilon/\varepsilon_0$ is temperature independent for both the incipient FM and the incipient AF. This immediately explains observation (i) above.

As T is decreased $\langle M_q M_{-q} \rangle$ develops q dependence, peaking near q = 0 for the FM case and near the magnetic Bragg-vector at the zone boundary for the AF case. To determine the temperature dependence of ε, we need a microscopic theory for g(q). The dielectric constant depends on the long wave-length longitudinal and transverse optic phonon frequencies through the Lyddane-Sachs relation. We suppose that the microscopic origin of g(q), the coupling between the polarization and spin-correlations in Eq. (1), arises from the coupling of magnetic fluctuations to the optic phonon frequencies. We expand the exchange integral of neighboring spins on the distance between the atoms carrying the spins expanded in terms of the normal coordinates for these phonons, which is expressed as

$$H_{ex} = \sum_{i<j} J_{ij} S_i \cdot S_j \approx \sum \left[ J_{ij}(R_{ij}^0) + \vec{J}' \cdot (u_i - u_j) + J''(u_i - u_j)^2 + \ldots \right] (S_i \cdot S_j) . \qquad (5)$$

The first term in Eq. (5), proportional to $(u_i\text{-}u_j)$, affects the phonon frequencies only to second-order and is therefore related to four-spin couplings. This will be less important than the term proportional to $(u_i\text{-}u_j)^2$ which changes frequencies of transverse and longitudinal polarized phonons in leading order proportional to $\langle S_i S_j \rangle$. Expanding $(u_i\text{-}u_j)^2$ in terms of the phonon coordinates $u_q$ and, keeping only the long wavelength modes relevant for determining the dielectric constant ($q \to 0$), we find the coupling has the form:

$$J'' u_0^2 \sum_q (1 - \cos q \cdot R_{ij}^0) \langle S_q S_{-q} \rangle \qquad (6)$$

where $R_{ij}^0$ are the nearest neighbor coordinates. Then, using our assumption that the shift in dielectric constant can be cast in terms of a frequency shift in the optic phonons, we find that

g(q) in Eq. (3) is proportional to $(1-\cos q\cdot R_{ij}^{0})$. For ferromagnets, this coupling vanishes as $q^2$ in the long wavelength limit where $\langle M_q M_{-q}\rangle$ develops a peak as T approaches $T_c$. For antiferromagnets, the coupling is a maximum for q near a zone-boundary where $\langle M_q M_{-q}\rangle$ develops a peak as T approaches $T_N$ [15].

We can now qualitatively explain the observations (ii) to (iv) using Eq. (3) and such a form for g(q). This type of problem has been examined for investigating the change in resistivity near a FM or AF transition [16,17] and the effects of AF fluctuations on s-wave superconductivity [18]. As T is decreased towards $T_c$, $\langle M_q M_{-q}\rangle$ develops a peak around q = 0 with a width proportional to the correlation length ξ(T). But in this region, the contribution to the integrand of I(T) is suppressed by a factor of $q^4$. The major part of the integrand for I(T) remains unchanged except very close to the transition where $R_{ij}^{0}/\xi << 1$, and a reduction (enhancement) in ε is expected from (4) for J">0 (J"<0). Our experiments are not precise enough to reveal this critical region.

On the other hand, for the incipient AF, as T is decreased $\langle M_q M_{-q}\rangle$ develops an increasing peak in the region where g(q) is nearly a constant. This is also the region of most of the phase space in the integral of I(T). Therefore a larger effect in the AF is to be expect, as found. From the experiment, an increase in ε corresponds to J"<0. In the critical regime ξa>>1, ie for $|T-T_N|/T_N << 1$, a peak in ε similar to the specific heat is to be expected following the theory of Fisher and Langer [16]. Some hint of a peak may be found in Fig. (1).

We turn now to the region T<<$T_c$ or $T_N$. For the AF, as well as the FM, in the classical approximation $\langle M_q M_{-q}\rangle$ ~$M^2$, where M is the staggered moment at $q = 2\pi/R_{ij}^{0}$ for the AF and at

q=0 for the FM. We see immediately using the derived form for g(q) that $I(T) \rightarrow 0$ in this regime for the FM. With J"<0, this means that for the FM $\varepsilon \sim \varepsilon_0$ for $T<<T_c$ which is lower than the asymptotic value for $T>>T_c$. More generally, in ferromagnets exhibiting magnetodielectric behavior, the intrinsic dielectric constant $\varepsilon_0$ is only measured at low temperature, where the magnetic fluctuations are frozen out. On the other hand, for the AF, there is essentially no change in I(T) for $T<<T_N$ compared to $T\sim T_N$ (excepting the critical regime). The contrasting behaviors in ε(T) for the AF and FM are thus qualitatively explained. We can further test these ideas in the FM, where according to the argument above, the decrease in the dielectric constant for $T<T_c$ is simply proportional to the ordered $M^2$. This is shown in Fig. (2) at both zero field and at finite field.

Finally we discuss observation (iv) regarding the different magnetic field dependence for $T\sim T_c$ and for $T\sim T_N$. A sharp effect near $T\sim T_c$ is to be expected, since near $T_c$, M is a strongly non-linear function of H due to the switching of domains. This is exhibited in Fig. (3) where the relative change in ε vs H is plotted together with the measured $M^2$. Only smooth behavior is to be expected for small H near $T_N$ in the AF. For large fields, a uniform magnetization does develop in the AF, so effects similar in magnitude to that in the FM are to be expected as found in Fig. (3).

We have investigated the temperature and magnetic field dependence of the dielectric constant for both $SeCuO_3$ (FM) and $TeCuO_3$ (AF) and have developed a simple model for understanding how the q dependent spin-spin correlations change the measured capacitance. By positing that the MD effect arises from long wavelength frequency shifts in the optical phonons induced by magnetic fluctuations, we find an expression for the microscopic q-dependent coupling between uniform polarization and the spin-spin correlation function. This model gives good qualitative agreement with the experimental measurements of ε — both the temperature dependence of the dielectric constant and the magnetocapacitance can be expressed simply in terms of the temperature and field variations of magnetic fluctuations and the uniform magnetization.

The practical interest in understanding magnetoelectric (and now magnetodielectric) couplings arises from device applications. The analysis presented above, based on a comparison of the MD response of FM $SeCuO_3$ and AF $TeCuO_3$, suggests routes to materials with higher MD coefficients. The data and our model show that antiferromagnets might demonstrate the larger overall temperature effect on ε through the q-dependence of g(q), especially at temperatures just above $T_N$. However, for effective coupling to an external magnetic field, a ferromagnetic order parameter would be required. It is conceivable that these two ingredients could be engineered into a material using either thin film deposition or nanoscale synthesis techniques. Here, the FM and AF components should have spatial proximity and be strongly coupled. Since there are few insulating ferromagnets, such nanostructuring could also provide a way to ensure the composite material is an effective dielectric. If such a material were feasible, it would resemble a dielectric version of exchange-coupling, which provides a basis for giant magnetoresistance.

**Figures**

Figure 1. Magnetization of $SeCuO_3$ versus temperature in 1 kOe field (upper) and dielectic constant of $SeCuO_3$ versus temperature in zero applied field (lower). The inset plots $\varepsilon(T)$ near $T_c$ for different applied fields.

Figure 2. Comparison of $\varepsilon$ to $M^2$; the agreement is predicted by Eq. 4.

Figure 3: Suppression of $\varepsilon$ as a function of magnetic field at $T_c$ ($T_N$) in $SeCuO_3$ ($TeCuO_3$). Note that the bottom plot showing $\varepsilon(H)$ for $SeCuO_3$, plots the magnetic field on a different scale than the upper two plots. The solid lines show the measured values of $M(H)^2$.

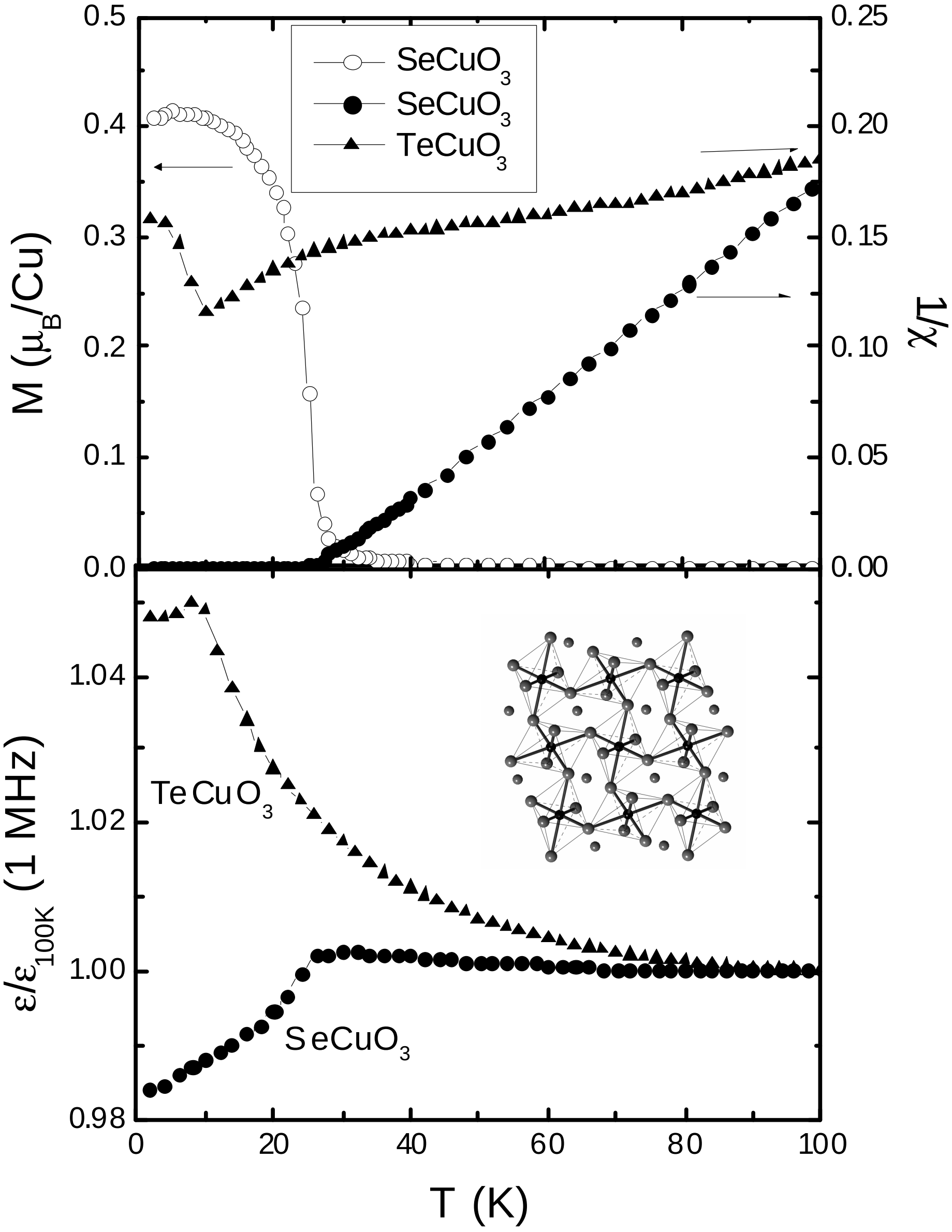

FIGURE 1

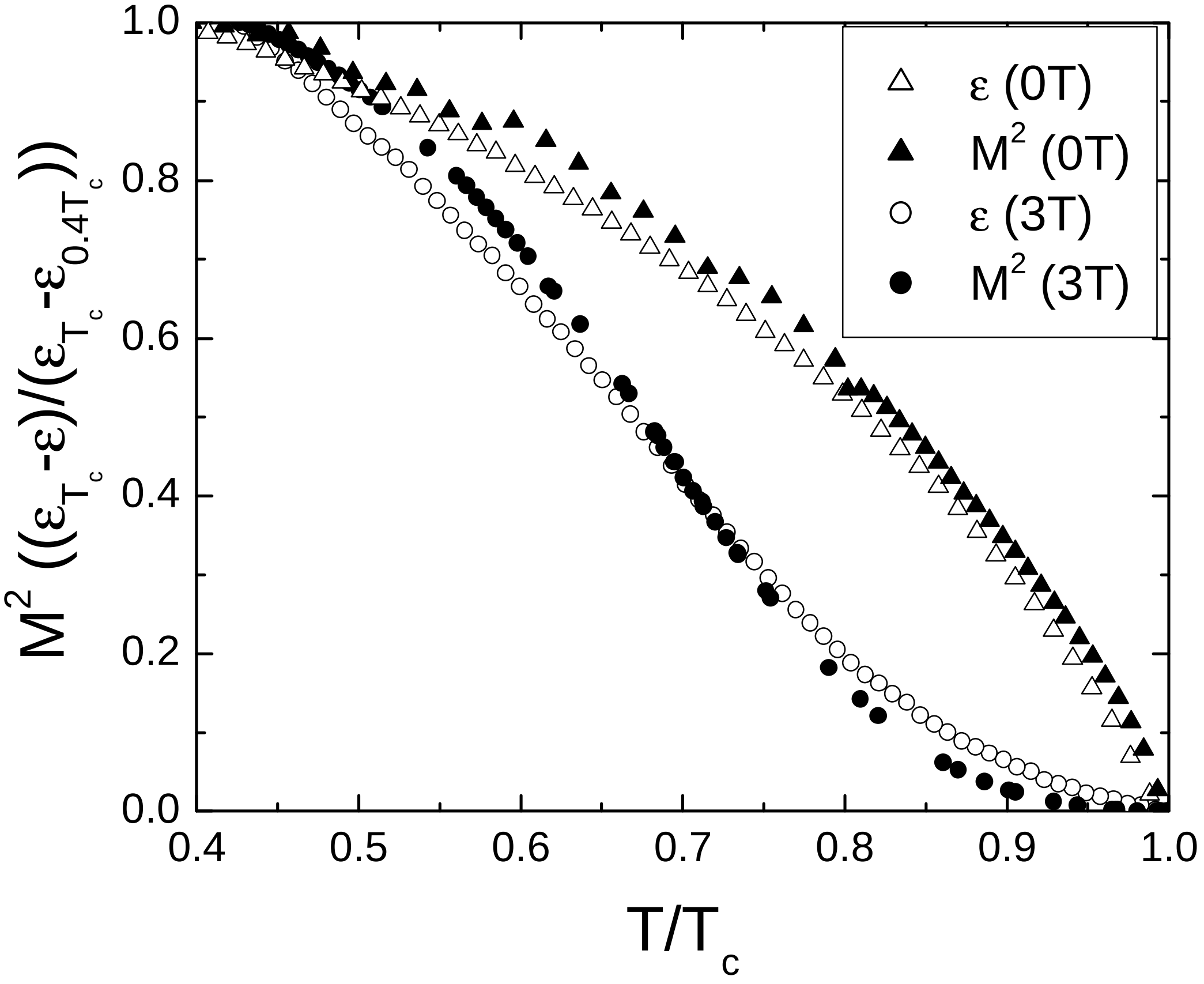

FIGURE 2

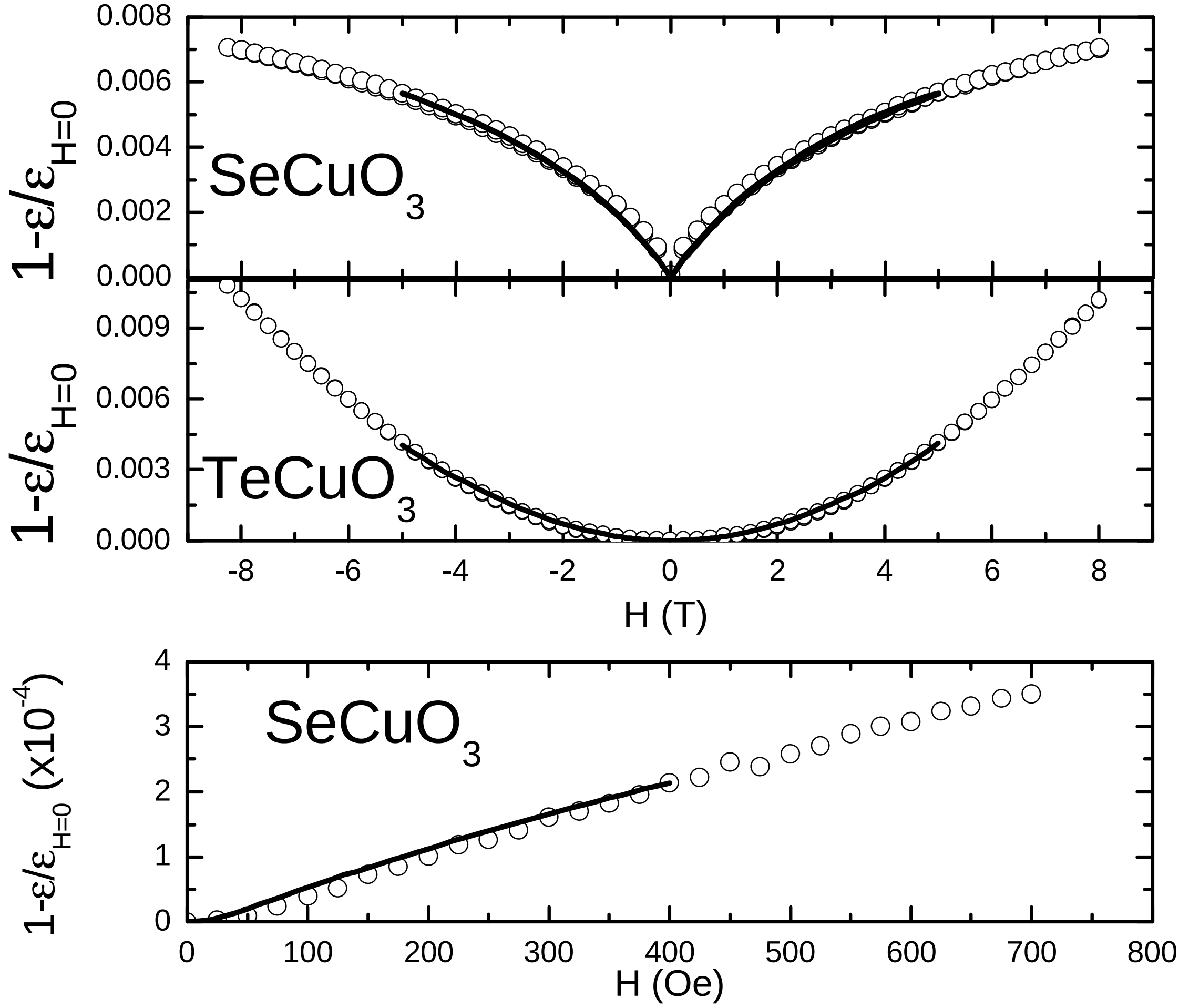

FIGURE 3